\documentclass[pra,twocolumn,showpacs,superscriptaddress]{revtex4}

\usepackage{amsmath}
\usepackage{graphicx}
\usepackage{dcolumn}
\usepackage{bm}

\begin{document}

\preprint{ITP/UU-XXX}

\title{Interacting Preformed Cooper Pairs in Resonant Fermi Gases}

\author{K. B. Gubbels}

\email{K.B.Gubbels@uu.nl}

\affiliation{Fritz-Haber-Institut der Max-Planck-Gesellschaft, Faradayweg 4-6, D-14195 Berlin,\\ Germany}

\affiliation{
Radboud University Nijmegen, Institute for Molecules and Materials,
Heyendaalseweg 135, 6525 AJ Nijmegen, \\
The Netherlands}

\affiliation{
Institute for Theoretical Physics, Utrecht University,
Leuvenlaan 4, 3584 CE Utrecht,\\ The Netherlands}

\author{H. T. C. Stoof}

\affiliation{
Institute for Theoretical Physics, Utrecht University, Leuvenlaan 4, 3584 CE Utrecht,\\ The Netherlands}


\begin{abstract}

We consider the normal phase of a strongly interacting Fermi gas, which can have either an equal or an unequal number of atoms in its two accessible spin states. Due to the unitarity-limited attractive interaction between particles with different spin, noncondensed Cooper pairs are formed. The starting point in treating preformed pairs is the Nozi\`{e}res-Schmitt-Rink (NSR) theory, which approximates the pairs as being noninteracting. Here, we consider the effects of the interactions between the Cooper pairs in a Wilsonian renormalization-group scheme. Starting from the exact bosonic action for the pairs, we calculate the Cooper-pair self-energy by combining the NSR formalism with the Wilsonian approach. We compare our findings with the recent experiments by Harikoshi {\it et al.} [Science {\bf 327}, 442 (2010)] and Nascimb\`{e}ne {\it et al.} [Nature {\bf 463}, 1057 (2010)], and find very good agreement. We also make predictions for the population-imbalanced case, that can be tested in experiments.
\end{abstract}


\maketitle

\section{Introduction}

Pair formation is a fundamental process in nature. In particular, it is the underlying mechanism for superconductivity and superfluidity in interacting fermionic systems. The impressive amount of control currently achievable in experiments with ultracold atomic quantum gases makes these systems ideally suited for detailed studies of strong correlations in many-body systems. Examples of such control include the cooling of the gases down to the nanoKelvin regime, the tuning of the interatomic interaction strength by means of Feshbach resonances, the manipulation of the number of atoms in a particular quantum state, and the shaping of the confining potential \cite{UFG,revBDZ,SGD,Grimmrev}. Due to their unique properties, ultracold atomic gases are sometimes referred to as `ideal quantum simulators'.

By varying the interaction strength between fermionic atoms in a different internal state, it is possible to perform detailed experimental studies of the continuous crossover between a Bardeen-Cooper-Schrieffer (BCS) superfluid of loosely bound Cooper pairs and a Bose-Einstein condensate (BEC) of tightly bound molecules \cite{cross1,cross2,cross3,cross4,cross5,cross6}, which results in a unified view of these two apparently different limits of superfluidity. In the intermediate regime of the crossover, where the scattering length of the interaction diverges, a novel superfluid was realized. In this so-called unitarity limit, the size of the Cooper pairs is comparable to the interparticle distance and the superfluid has remarkable universal properties \cite{Houni,Tanrelation1,ThomasExpEnt,UedaExp,SalomonExpNat}.

An early theoretical exploration of the BEC-BCS crossover by Leggett was performed at zero temperature and made use of the mean-field BCS Ansatz \cite{crossleg}. The BCS theory takes into account a condensate of pairs, but ignores the noncondensed pairs. Therefore, it is not even suitable on a qualitative level to describe the critical temperature curve of the crossover, since at the BEC side superfluidity is lost due to pairs being thermally excited into nonzero momentum states.

The Nozi\`{e}res-Schmitt-Rink (NSR) theory for the normal state of a balanced Fermi gas \cite{NSRpaper}, which takes into account a noninteracting gas of noncondensed Cooper pairs, improves the BCS theory significantly and provides a remarkably good description of the critical temperature $T_{\rm c}$ for all interaction strengths \cite{Melo1993}. The NSR theory does not change the linearized gap equation for the superfluid transition temperature, so that the relationship between the (average) fermionic chemical potential $\mu$ and $T_{\rm c}$ is unchanged with respect to BCS theory. However, the NSR theory does change the equation of state, which affects the total particle density $n(\mu,T)$. As a result, the ratio $k_{\rm B}T_{\rm c}/\epsilon_{\rm F}$ is altered, with the Fermi energy given by $\epsilon_{\rm F}=\hbar^2(3\pi^2 n)^{2/3}/2 m$ and $m$ the mass of the fermions. At unitarity, NSR theory predicts rather large values for the ratios $k_{\rm B}T_{\rm c}/\mu$ and $\epsilon_{\rm F}/\mu$ compared to Monte-Carlo (MC) results. Nevertheless, the NSR prediction $k_{\rm B}T_{\rm c}=0.23 \epsilon_{\rm F}$ is still quite close to the MC result of $k_{\rm B}T_{\rm c}=0.15 \epsilon_{\rm F}$ \cite{Burovski}. It was also shown that away from the critical temperature the NSR theory gives excellent agreement with accurate thermodynamic measurements on the unitary balanced Fermi gas \cite{ThomasExpEnt,UedaExp,SalomonExpNat,Hu,Hu2}.

In the beginning of 2006, two experimental groups performed the first experiments on an ultracold atomic Fermi gas with a population imbalance in its two accessible spin states \cite{KetterleImb,HuletImb}. The phase diagram was experimentally shown to be dominated by a tricritical point \cite{ShinImb}, that was predicted from mean-field theory, NSR theory and renormalization-group theory \cite{GubbelsImbMF,Parish,GubbelsImbRG}. However, in Ref.~\cite{Parish} it was already noticed that the NSR theory breaks down for small spin imbalances. Namely, near the critical temperature, the NSR theory predicts a negative polarization $p=(n_{+}-n_{-})/(n_{+}-n_{-})$ for a positive chemical potential difference $(\mu_{+}-\mu_{-})$, which corresponds to a compressibility matrix $-\partial^2 \omega/\partial \mu_{\sigma} \partial \mu_{\sigma'}$ that is not positive definite \cite{Parish}. Here, $n_{\pm}$ are the atomic densities for the two spin states $\sigma = \pm$, $\mu_{\sigma}$ is the chemical potential for spin state $\sigma$ and $\omega$ is the thermodynamic potential density of the Fermi gas. It is a quite unsatisfactory situation that the NSR theory, which gives such a good agreement with accurate thermodynamic experiments for the strongly-interacting normal state of the Fermi gas \cite{UedaExp,SalomonExpNat,Hu2}, already gives unphysical results for even the smallest population imbalances.

For this reason, we improve the theory of Cooper-pairs by taking also into account the effect of the interactions between the noncondensed pairs. If the microscopic fermionic action is exactly transformed into a Cooper-pair action by means of the so-called Hubbard-Stratonovich transformation \cite{stratonovich,hubbard2}, then the resulting action not only contains a noninteracting part, but also two-pair interactions, three-pair interactions, and all higher-order interactions. In the BEC limit of the crossover, the tightly bound pairs interact repulsively with a scattering length given by $0.6 a$ with $a$ the scattering length of the interaction between the fermions \cite{Gora06a}. A Popov theory for the composite bosonic pairs that includes the pair interaction effects was formulated by Pieri and Strinati, leading in the BEC regime to Popov's results for point-like bosons  \cite{PieriBEClim,PieriPopov}. Below the critical temperature, also the Bogoliubov theory for interacting Cooper pairs was studied \cite{Romans,Hu,Diener}. Other strong-coupling approaches that go beyond the NSR theory include so-called self-consistent ladder approximations  \cite{Haussmann, ZwergerCross} and Monte-Carlo calculations \cite{CarlsonUni,Astrakharchik,Burovski,BulgacUni,TroyerCross}. It is somewhat surprising that these calculations do not seem to lead to a better agreement with most recent experiments on the equation of state for a unitary Fermi gas than the NSR theory \cite{SalomonExpNat,Hu2}.

Viewing the normal state of a strongly interacting Fermi mixture as a gas of interacting Cooper pairs is complementary to a Fermi-liquid picture \cite{SalomonFL}. The Fermi-liquid picture focusses on the single-particle correlation function, or fermionic Green's function, while the Cooper-liquid picture focusses on the two-particle correlation function, or the pair Green's function. Possible differences between the two pictures arise only from different approximation schemes. The additional advantage of the Cooper-liquid picture is that it gives a clear condition for the transition to the superfluid phase, namely when the (effective or renormalized) chemical potential of the Cooper-pairs goes to zero.

To study interaction effects in an interacting Bose gas in a nonperturbative manner, the renormalization group (RG) approach is an established approach \cite{wilson}. RG studies for interacting bosons on the verge of becoming superfluid have increased our understanding of the resulting phase transition \cite{rgfisher,rgpaper,KopietzBosTc,BlaizotTc,FloerchingerBEC}. In this article, we perform a renormalization-group study of interacting Cooper pairs in the unitarity limit. To this end, we generalize the Wilsonian RG theory for point-like bosons to the more complicated case of Cooper pairs. The article is organized as follows. First, we briefly discuss the exact action for the Cooper pairs that can be derived from the microscopic fermionic action. Then, we set up the Wilsonian renormalization scheme for Cooper pairs in order to calculate the effects of the Cooper-pair interactions. In particular, we calculate the self-energy of the Cooper pairs. We also compare our results with the recent experiments by Harikoshi {\it et al.} \cite{UedaExp} and Nascimb\`{e}ne {\it et al.} \cite{SalomonExpNat}. The agreement with these detailed experiments on the equation of state of the unitary Fermi gas turns out to be very good. We also make predictions for the equation of state of the imbalanced gas, which can be experimentally tested in the near future.

\section{Exact action for Cooper pairs}\label{parac}

We start from the microscopic action for an ultracold two-component Fermi gas with a local interaction
\begin{eqnarray}\label{eqacfer}
S[\phi^*,\phi] \nonumber  &=& - \sum_{\sigma = \pm}
\int d \boldsymbol{\xi}d\boldsymbol{\xi}'~ \phi^*_{\sigma}(\boldsymbol{\xi})\hbar G^{0 \, -1}_{\sigma}(\boldsymbol{\xi},\boldsymbol{\xi}') \phi_{\sigma}(\boldsymbol{\xi}') \nonumber \\
&&+V_{0} \int d \boldsymbol{\xi}~  
\phi^*_{+}(\boldsymbol{\xi}  ) \phi^*_{-}(\boldsymbol{\xi}  )
\phi_{-}(\boldsymbol{\xi}  ) \phi_{+}(\boldsymbol{\xi}  ),
\end{eqnarray}
where $ \phi_{\sigma} (\boldsymbol{\xi}) $ is the fermionic field associated with the annihilation of a particle with spin $\sigma$ at $\boldsymbol{\xi} =(\tau,{\bf x})$. Here, ${\bf x}$ denotes the position and $\tau$ the imaginary time. Moreover, $V_0$ is the strength of the local interaction, while the Fourier transform of the inverse noninteracting Green's function $G^{0 \, -1}_{\sigma}$ is given by $\hbar G^{0 \, -1}_{\sigma,n,{\bf k}}=i\hbar\omega_{n}-\epsilon_{\bf k}+\mu_{\sigma}$, with $\mu_{\sigma}$ the fermionic chemical potential for spin state $\sigma$, $\epsilon_{\bf k}=\hbar^2{\bf k}^2/2m$ the kinetic energy, and $\omega_{n}$ the odd fermionic Matsubara frequencies. Note that in Eq.~(\ref{eqacfer}) the integration over $\tau$ is from 0 to $\hbar/ k_{\rm B}T$ with $T$ the temperature. For atomic gases the spin label $\sigma$ refers to two hyperfine states that are used in the experiment to realize the two-component gas. The action of Eq.~(\ref{eqacfer}) was previously used as the starting point for a RG study of the population-imbalanced Fermi gas in its normal phase \cite{GubbelsImbRG}.

For an attractive interaction, the purely fermionic microscopic action can be exactly transformed into a Bose-Fermi action that contains the pairing field $\Delta({\bf x},\tau)$ by means of the Hubbard-Stratonovich transformation. For the precise procedure, see e.g. Ref.~\cite{SGD}. The result is
\begin{eqnarray}\label{eqacbosfer}
&&S[\phi^*,\phi,\Delta^*, \Delta]   = \\ 
&&-\sum_{\sigma = \pm}
\int d \boldsymbol{\xi}d\boldsymbol{\xi}'~ \phi^*_{\sigma}(\boldsymbol{\xi})\hbar G^{0 \, -1}_{\sigma}(\boldsymbol{\xi},\boldsymbol{\xi}') \phi_{\sigma}(\boldsymbol{\xi}') \nonumber \\
&&+ \int d \boldsymbol{\xi}\left\{ -
\frac{|\Delta(\boldsymbol{\xi})|^2}{V_{\bf 0}}+\Delta^*(\boldsymbol{\xi})\phi_{+}(\boldsymbol{\xi}  ) \phi_{-}(\boldsymbol{\xi}  )+{\rm c.c.}\right\},\nonumber
\end{eqnarray}
where the last two terms of the second line indeed show that two fermions of opposite spin can form a pair, or that a pair can decay into two fermions. The action of Eq.~(\ref{eqacbosfer}) can be interpreted as an interacting Bose-Fermi mixture \cite{WolfgangBosFerMix}, and has been the starting point of the functional renormalization group studies in Refs.~\cite{Birse,Diehl}. 

Here, we follow a different route. We start by integrating out the fermions exactly, resulting in an exact microscopic action for the Cooper pairs. We call this action for the pairs microscopic, because no pair fluctuation effects have been taken into account yet. Incorporating these fluctuations exactly would consequently result in the exact effective action for the pairs. The microscopic Cooper pair action has a very rich structure in momentum and frequency space, i.e., a highly nonlocal character, which in particular is true for the obtained Cooper-pair vertices.  In this study, we take the complicated structure of the microscopic Cooper-pair propagator fully into account when performing the RG calculations. However, when calculating the effective pair propagator with our RG, we only consider the renormalization of the momentum- and frequency-independent part of the pair propagator, i.e., we do not consider the renormalization of the pair effective mass. After the exact integration over the fermionic fields, Eq.~(\ref{eqacbosfer}) results in (see again e.g. Ref.~\cite{SGD})
\begin{eqnarray}\label{eqacbos}
&&S[\Delta^*, \Delta]   =  -\int d \boldsymbol{\xi}
\frac{|\Delta(\boldsymbol{\xi})|^2}{V_{\bf 0} } -\hbar{\rm Tr} \log (-{\bf G}^{-1}),
\end{eqnarray}
where the Nambu space inverse Green's function ${\bf G}^{-1}$ is given by $G_{11}^{-1}= G^{0 \, -1}_{+}(\boldsymbol{\xi},\boldsymbol{\xi}')$, $G_{22}^{-1}=-G^{0 \, -1}_{-}(\boldsymbol{\xi}',\boldsymbol{\xi})$, $\hbar G_{12}^{-1}=-\Delta(\boldsymbol{\xi})\delta(\boldsymbol{\xi}-\boldsymbol{\xi}')$ and  $G_{21}^{-1}=G_{12}^{-1\,*}$. By expanding the logarithm in Eq.~(\ref{eqacbos}) in powers of $\Delta$, we obtain an infinite series that prohibits an exact solution to the problem, so that approximations have to be made in order to proceed. By performing the mean-field, or saddle-point, approximation, the full path integral over the bosonic field $\Delta(\boldsymbol{\xi})$ is simply approximated by the value of the integrand associated with the global minimum $S[\Delta_0^*,\Delta_0]$. This approximation results in the well-known BCS thermodynamic potential. 

Going beyond mean-field theory, the next step is to perform a Gaussian, or random-phase approximation. In the normal phase, this means that Eq.~(\ref{eqacbosfer}) is expanded up to second order in the pairing field, and that the resulting Gaussian functional integral is performed exactly. By neglecting the higher-order contributions, the Cooper pairs are thus physically approximated as forming a noninteracting gas. The resulting theory is also called the Nozi\`{e}res-Schmitt-Rink approximation \cite{NSRpaper}, and it has been applied with success to the study of thermodynamic properties above and below the critical temperature \cite{Romans,Hu,Diener}. It is namely possible to generalize this theory also to the superfluid state by making in Eq.~(\ref{eqacbosfer}) the substitution $\Delta(\boldsymbol{\xi})=\Delta_0+\Delta'(\boldsymbol{\xi})$, expanding Eq.~(\ref{eqacbos}) up to second order in the fluctuations $\Delta'$, and performing the resulting Gaussian functional integral. 

\begin{figure}
\begin{center}
\includegraphics[width=1.0\columnwidth]{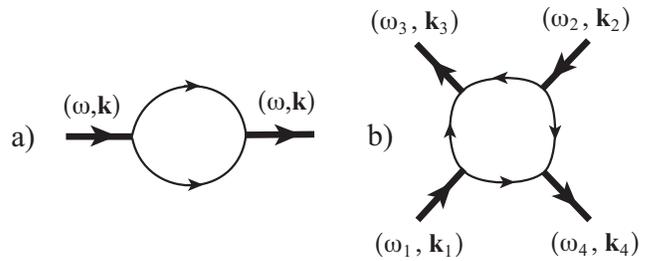}
\caption{\label{figbare} Diagrammatic representation of a) the bare Cooper-pair propagator and b) the bare Cooper-pair interaction. The Cooper pairs are represented by thick lines, while the thin lines correspond to fermionic propagators. }
\end{center}
\end{figure}

In this article, we only study the normal phase. The inverse propagator for the noncondensed Cooper pairs $G_{\Delta}^{{\rm B}~-1}$ that follows from the quadratic part in the pairing field of Eq.~(\ref{eqacbos}), is given by (see e.g. Ref.~\cite{SGD})
\begin{eqnarray}\label{eqprop}
&& \hbar G_{\Delta}^{{\rm B}~-1}(i\omega_n, k)=\\
&&\frac{m}{4 \pi \hbar^2 a}+ \frac{1}{V}\sum_{\bf k'} \left\{\frac{1-f_+(\epsilon_{\bf k'})-f_{-}(\epsilon_{\bf k-k'})}{-i\hbar\omega_n+\epsilon_{\bf k'}+\epsilon_{\bf k-k'}-2\mu}-\frac{1}{2\epsilon_{\bf k'}}\right\},\nonumber
\end{eqnarray}
with $V$ the volume, $a$ the scattering length, $\mu=(\mu_{+}+\mu_{-})/2$ the avarage chemical potential, $f_{\sigma}(\epsilon)=1/\{e^{\beta(\epsilon-\mu_{\sigma})}+1\}$ the Fermi distribution function, $\beta=1/k_{\rm B} T$, and $\omega_n$ an even bosonic Matsubara frequency. The Feynman diagram that corresponds to the Cooper-pair propagator is shown in Fig.~\ref{figbare}a. We call $G_{\Delta}^{{\rm B}~-1}$ the bare or microscopic propagator, indicating that no Cooper-pair interaction effects have been taken into account yet. Note that the bare propagator is exact, in the sense that it follows from an exact transformation of the fermionic action.  With our RG approach we can consequently systematically include Cooper-pair interaction effects that lead to self-energy corrections to the bare propagator.

The Cooper-pair interaction $V^{\rm B}_{\Delta}$ follows from the quartic part in the pairing field of Eq.~(\ref{eqacbos}) and is diagrammatically represented in Fig.~\ref{figbare}b \cite{PieriBEClim,PieriPopov}. Here, we do not take the full frequency and momentum dependence of the Cooper-pair interaction vertex into account, but we consider only two external frequencies and momenta to be nonzero, namely either $\omega_1=-\omega_2$ and ${\bf k}_1=-{\bf k}_2$, or $\omega_3=-\omega_4$ and ${\bf k}_3=-{\bf k}_4$, where the labeling is given in Fig.~\ref{figbare}b. This specific choice corresponds physically to considering only zero center-of-mass frequencies and momenta, which is motivated in the next section. The resulting expression is given by
\begin{eqnarray}\label{eqint}
&& V^{\rm B}_{\Delta}(i\omega_n, k) \equiv V^{\rm B}_{\Delta}G_{V}(i\omega_n, k) =\frac{1}{\hbar^3\beta V} \times\\
&& \sum_{n',{\bf k'}}G^0_{-,n',{\bf k'}} G^0_{+,-n',-{\bf k'}}
 G^0_{-,n'+n,{\bf k'}+{\bf k}} G^0_{+,-n'-n,-{\bf k'}-{\bf k}},\nonumber
\end{eqnarray}
where we have defined $V^{\rm B}_{\Delta}\equiv V^{\rm B}_{\Delta}(0,0)$, so that $G_V$ encapsulates the considered (relative) momentum and frequency dependence of the Cooper-pair interaction. The Matsubara sum over odd fermionic frequencies $n'$ in Eq.~(\ref{eqint}) is readily performed analytically, but results in a somewhat cumbersome expression. We call $V^{\rm B}_{\Delta}(i\omega_n,k)$ the bare or microscopic interaction, in order to make the distinction with the effective or renormalized Cooper-pair interaction, which includes the effect of Cooper-pair fluctuations and that is calculated in the next section during the RG flow.

Due to the repulsive interaction between the Cooper pairs, they acquire a self-energy $\Sigma_{\Delta}$, of which the momentum- and frequency-independent part is most relevant for the RG flow. So the full propagator becomes $G^{-1}_{\Delta}(i\omega_n,k)=G_{\Delta}^{{\rm B}~-1}(i\omega_n,k)-\Sigma_{\Delta}$. We define the bare Cooper-pair chemical potential as $\mu^{\rm B}_{\Delta}\equiv \hbar G^{{\rm B}~-1}_{\Delta}(0,0)$, while the renormalized chemical potential is given by $\mu_{\Delta}=\mu^{\rm B}_{\Delta}-\hbar\Sigma_{\Delta}$. The renormalized chemical potential thus includes the self-energy effects. As a result, the full Cooper-pair propagator $G_{\Delta}$ depends on the renormalized chemical potential $\mu_{\Delta}$. In this study, we do not take the frequency and momentum dependence of the Cooper-pair self-energy into account. Doing this would result also in a renormalization of the effective mass of the Cooper pairs, an interesting topic for further study. Note that the Cooper-pairing fields do not have the same units as the fields for point-like bosons, resulting also in a different unit for the propagator. Therefore, the present definitions of the Cooper-pair chemical potential and the self-energy do not have the unit of energy. However, we still think our nomenclature is appropriate, due to the physical and mathematical analogy with the corresponding concepts for point-like bosons.

A complementary physical meaning for the chemical potential of the Cooper pairs is obtained by realizing that the noncondensed Cooper pairs mediate an interaction between the fermions as follows from the action in Eq.~(\ref{eqacbosfer}). The bare Cooper pair propagator of Eq. (\ref{eqprop}) is indeed equivalent to the many-body transition matrix for the fermions in the ladder-diagram approximation, which we call the bare many-body transition matrix. At zero energy and momentum, we have for the bare many-body scattering length $a^{{\rm B}}_{\rm MB}$ that $\hbar G^{\rm B}_{\Delta}(0,0)= 1/\mu^{\rm B}_{\Delta}=T^{\rm B}_{\rm MB}(0,0) = 4 \pi \hbar^2a^{\rm B}_{\rm MB}/m$. The renormalized chemical potential for the Cooper pairs then corresponds to a renormalized transition matrix for the fermions that includes more Feynman diagrams than the bare one, for which the renormalized many-body scattering length  $a_{\rm MB}$ is given by $\mu_{\Delta}=m/ 4 \pi \hbar^2a_{\rm MB}$.

\section{Renormalization formalism}

To treat the interaction effects of the Cooper pairs in a nonperturbative manner we use the Wilsonian renormalization group approach \cite{wilson}. The procedure goes as follows. First an integration is performed over degrees of freedom in a high momentum shell of infinitesimal width. The result of this integration is consequently absorbed into various coupling constants of the theory, which are said to flow if degrees of freedom in subsequent momentum shells are integrated out. We could also perform another step, namely a rescaling of the momenta,
frequencies, and fields. This is convenient, if we would wish to treat universal properties of critical
phenomena such as critical exponents by looking at so-called RG fixed points \cite{wilson}. In this article, however,
we do not wish to calculate universal critical exponents, but rather quantities like the self-energy of the Cooper pairs, for which rescaling
is not particularly useful. The renormalization group then serves
as a nonperturbative method to iteratively solve a many-body
problem, rather than as a mapping between actions with the same high-momentum cutoff from which
critical scaling relations can be derived.

Thus, the first step of the method is to
evaluate the Feynman diagrams that renormalize the coupling
constants of interest, while keeping the integration over the
internal momenta restricted to the considered high-momentum shell. Only one-loop diagrams contribute to the
flow, because the width of the momentum shell is infinitesimally small
and each loop introduces an additional factor proportional to the infinitesimal width. Although the one-loop structure of the infinitesimal Wilsonian RG is exact, it does not mean that it is easy to also obtain exact results, since this would require the consideration of an infinite number of coupling constants. Although the latter is usually not possible in practice, the
RG distinguishes between the relevance of the coupling constants, so that a small set of them may
already lead to accurate results. 

\begin{figure}
\begin{center}
\includegraphics[width=1.0\columnwidth]{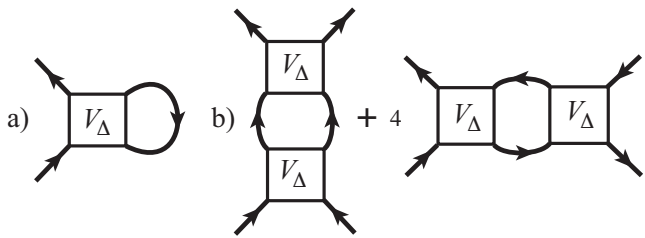}
\caption{\label{figbeta} Diagrammatic representation of the `$\beta$ functions'. a) Feynman diagram determing the self-energy of the Cooper pairs. b) Feynman diagrams renormalizing the Cooper-pair interaction. The middle diagram is also called the ladder diagram, the right diagram is called the bubble diagram. Note that the lines are thick and correspond to the Cooper-pair propagator.}
\end{center}
\end{figure}

The simplest RG calculation that gives nontrivial results treats the renormalization of the chemical potential $\mu_{\Delta}$ and the interaction strength $V_{\Delta}$. It ignores the three-pair interactions and higher. The flow equations for these two coupling constants can be derived along exacly the same lines as for point-like bosons, for which detailed accounts can be found in Refs. \cite{rgpaper,SGD}. The corresponding one-loop Feynman diagrams are diagrammatically represented in Fig. \ref{figbeta}. The main difference with point particles is that the frequency dependence of the Cooper-pair propagator is more complicated. As a result, the Matsubara sums of the one-loop Feynman diagrams cannot be performed analytically anymore, but have to be evaluated numerically within each momentum shell.

The RG flow of the Cooper-pair interaction strength and chemical potential are determined by the following set of coupled differential equations
\begin{eqnarray}\label{eqrg}
\frac{d\mu_{\Delta}}{d l} = \beta_{\mu}(l,\mu_{\Delta},V_{\Delta}),\quad \frac{dV_{\Delta}}{d l} = \beta_{V}(l,\mu_{\Delta},V_{\Delta})~,~
\end{eqnarray}
where the `$\beta$-functions' are given by
\begin{eqnarray}\label{eqbetamu}
\beta_{\mu} &=& -k_l'\frac{k_l^2V_{\Delta}}{\pi^2\beta}\sum_n \hbar G_{\Delta}(i\omega_n,k_l,\mu_{\Delta}),\\
\beta_{V} &=& k_l'\frac{k_l^2V^2_{\Delta}}{2\pi^2} \left\{\Xi(k_l,\mu_{\Delta})+4\Pi (k_l,\mu_{\Delta}) \right\}.~\label{eqbetav}
\end{eqnarray}
Here, $\Pi$ and $\Xi$ are the so-called `bubble' and `ladder' contributions to the effective Cooper-pair interaction, which are explained in the next paragraph. Moreover, $k_l$ denotes the wavevector of the Cooper pairs in the shell of infinitesimal width. This wavevector is parametrized by the flow parameter $l$, and we start the RG flow at the high-momentum cutoff $\hbar\Lambda$ and decrease as $k_l=\Lambda e^{-l}$. In addition, $k_l'$ is the derivative of $k$ with respect to $l$. Solving Eq.~(\ref{eqrg}) for increasing $l$ means that we are including the effect of pair fluctuations with lower and lower momenta, while due to the coupling of the differential equations we automatically generate an infinite number of Feynman diagrams, showing the nonperturbative nature of the RG. The initial conditions for  Eq.~(\ref{eqrg}) are $\mu_{\Delta}(l=0)=\mu^{\rm B}_{\Delta}=\hbar G^{{\rm B}~-1}_{\Delta}(0, 0)$ and $V_{\Delta}(l=0)=V^{\rm B}_{\Delta}$, which are calculated from Eqs.~(\ref{eqprop}) and (\ref{eqint}).

The one-loop expression for the renormalization of the chemical potential in Eq.~(\ref{eqbetamu}) that determines the self-energy of the Cooper pairs, has a clear physical meaning, since it is seen to be proportional to the renormalized pair interaction strength and to the density of Cooper pairs. The `bubble' diagram $ \Pi(k,\mu_{\Delta})$ describes the effect of particle-hole fluctuations on the effective Cooper-pair interaction, where these particles are now Cooper pairs. It is given by  
\begin{eqnarray}\label{eqbubble}
\Pi(k,\mu_{\Delta}) = \frac{\hbar^2}{\beta}\sum_n G_{\Delta}(i\omega_n,k,\mu_{\Delta})^2.
\end{eqnarray}
The `ladder diagram' describes the Bose-enhanced scattering of the bosonic Cooper pairs, given by
\begin{eqnarray}\label{eqladder}
\Xi(k,\mu_{\Delta}) = \frac{\hbar^2}{\beta}\sum_n  G_{V}(i\omega_n,k)^2| G_{\Delta}(i\omega_n,k,\mu_{\Delta})|^2,
\end{eqnarray}
where the momentum and frequency dependence of the interaction in Eq.~(\ref{eqint}), i.e. $G_V$, is seen to enter. This frequency and momentum dependence is important, since otherwise Eq.~(\ref{eqladder}) ultimately would lead to an ultraviolet divergence. This divergence physically arises from approximating the pair interaction as a point interaction, which is therefore insufficient. We also note that the self-energy diagram of Fig.~\ref{figbeta}a and Eq.~(\ref{eqbetamu}), and the bubble diagram of Fig.~\ref{figbeta}b  and Eq.~(\ref{eqbubble}) do not lead to divergencies. As a result, our present scheme for including the Cooper-pair interactions is the minimal choice for obtaining divergence-free, or equivalently, cutoff independent results. 

The structure of Eqs.~(\ref{eqrg}) to (\ref{eqbetav}) is analogous to that for the RG equations in the normal state of Ref.~\cite{rgpaper}, where point-like bosons are treated, but there are also a few differences. In that reference, a trivial scaling was performed on the chemical potential and the interaction strength, which amounts only to a direct rewriting of the differential equations. Second: there, also three-body interaction effects were considered, which are ignored here. Third: the present Cooper-pair propagator is more complicated than the atom propagator, so that the Matsubara sums in Eq.~(\ref{eqbetamu}) and (\ref{eqbetav}) cannot be done analytically. This difference can be avoided by appoximating the inverse Cooper-pair propagator as $G^{-1}_{\Delta}(i\omega_n,k,\mu_{\Delta}) = (i\hbar\omega_n -\hbar^2  {\bf k}^2/2 m_{\Delta} + \mu'_{\Delta})/\hbar Z_{\Delta}$, meaning that we would perform a low-energy and long-wavelength expansion with $m_{\Delta}$ the effective mass of the Cooper pairs and $\mu'_{\Delta}=\hbar Z_{\Delta}\mu_{\Delta}$ having the unit of energy. In this article, however, we take the full frequency and momentum dependence of the propagator into account, since this gives the most accurate results. 

The last difference with Ref.~\cite{rgpaper} involves the ladder diagram of Eq.~(\ref{eqladder}). Such ladder diagrams are known to lead to an ultraviolet divergence, if the interaction is approximated as a point interaction, i.e., when $G_V=1$. For interacting bosonic or fermionic atoms, the point interaction approximation is often used. The problem is then most easily solved by choosing a value for the microscopic interaction that cancels the divergence and that leads in the two-body limit of the theory to the scattering length that is obtained from experiments. For our present RG flow of the interacting Cooper pairs, this procedure is not possible, since the bare Cooper-pair interaction strength is calculated exactly from the transformed microscopic action, and is not a free parameter that we can try to match to some experimentally known quantity. As a result, we really need to go beyond the point interaction approximation and take into account the momentum and frequency dependence of the pair interaction in order to obtain physical results \cite{PieriBEClim}. 

\section{Results}

\begin{figure}
\begin{center}
\includegraphics[width=1.0\columnwidth]{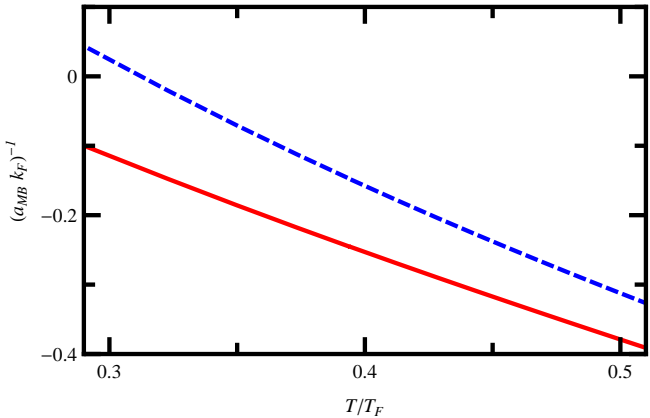}
\caption{\label{figselfen} (Color online) Bare many-body scattering length $a^{\rm B}_{\rm MB}$ (dashed line) and fully renormalized many-body scattering length $a^{\infty}_{\rm MB}$ (full line) as a function of temperature $T$ in the unitarity limit for the balanced Fermi gas. The Fermi wavevector $k_{\rm F}$ and Fermi temperature $T_{\rm F}$ are directly related to the total particle density through the Fermi energy $\epsilon_{\rm F}=k_{\rm B}T_{\rm F}=\hbar^2 k_{\rm F}^2/2m=\hbar^2(3\pi^2 n)^{2/3}/2 m$, which is calculated with the RG approach. The many-body scattering length is inversely proportional to the Cooper-pair chemical potential, and the difference between the bare and the renormalized value is due to the repulsive Cooper-pair interaction effects. }
\end{center}
\end{figure}

We compare our RG equations with recent detailed thermodynamic experiments that have measured the equation of state for a homogeneous unitary Fermi gas \cite{UedaExp,SalomonExpNat}. A relevant question is how our theory changes the results that follow from the NSR theory. As also found in previous studies, the results from the NSR approximation are in excellent agreement with experiments on the thermodynamics of the balanced Fermi gas at unitarity \cite{Hu2}. However, a fundamental problem with the NSR theory is that it gives already for small population imbalances rise to unphysical compressibilities  $-\partial^2 \omega/\partial \mu_{\sigma} \partial \mu_{\sigma'}$ \cite{Parish}. Moreover, the ratio of the critical temperature and the chemical potential is not so accurate for the NSR approximation, since it is the same as for mean-field theory, giving rise to $k_{\rm B}T^{\rm c}_{\rm MF} = 0.66 \mu$ for the balanced Fermi gas. This follows from the observation that when $G^{\rm B ~ -1}_{\Delta} (0,0) = \mu^{\rm B}_{\Delta} = m/4\pi \hbar^2 a^{\rm B}_{\rm MB}=0$, then both mean-field theory and NSR theory predict a transition to the superfluid state, i.e., the condensation of Cooper pairs. The result deviates more than a factor of two from the Monte-Carlo result  $k_{\rm B}T^{\rm c}_{\rm MC} = 0.31 \mu$ \cite{Burovski}.  So a theory that goes beyond NSR ideally should not have a large effect on the equation of state for the balanced case, should significantly lower the critical temperature, and give rise to physical results in the imbalanced case.

Studying the thermodynamics of the unitary Fermi gas in the grand-canonical ensemble requires the knowledge of the thermodynamic potential. The NSR theory of the strongly interacting normal state gives rise to two contributions to the thermodynamic potential, namely a contribution describing an ideal gas of fermions and a contribution describing an ideal gas of noncondensed Cooper pairs \cite{NSRpaper}. The ideal Fermi gas contribution follows directly from Eq.~(\ref{eqacbos}) as the part that is independent of the Cooper-pair field $\Delta$, and is therefore also not renormalized. It is given by
\begin{equation}
\omega_{\rm ig}(T,\mu_{\sigma}) = -\frac{1}{\beta V}\sum_{\bf k,\sigma} \log[1+e^{-\beta(\epsilon_{\bf k}-\mu_{\sigma})}].
\end{equation}
The contribution to the thermodynamic potential density due to the Cooper pairs is given by the one-loop expression
\begin{eqnarray}\label{eqtp}
\frac{d\omega_{\Delta}}{d l} = -k'_l\frac{k_l^2}{2\pi^2\beta} \sum_n \log[-G^{-1}_{\Delta}(i\omega_n,k_l,\mu_{\Delta})],
\end{eqnarray}
where the first minus sign on the right-hand side is only present when $k_l$ is a decreasing function. Note that this last expression gives precisely the differential form of the NSR contribution to the thermodynamic potential density when the Cooper-pair chemical potential is not renormalized ($\mu_{\Delta}\equiv\mu^{\rm B}_{\Delta}$), i.e., when we consider the Cooper pairs to be noninteracting. To evaluate Eqs.~(\ref{eqrg}) and (\ref{eqtp}) numerically, it is convenient to perform contour integration, leading to the results in Eqs.~(\ref{eqgfint}), (\ref{eqbubint}) and (\ref{eqtpint}) of the appendix \cite{NSRpaper}. If the exact Cooper-pair propagator is inserted in Eq.~(\ref{eqtp}), then the exact thermodynamic potential density is obtained. However, this would require the treatment of all $n$-body interactions, which is presently out of reach.

In this article, we have studied the effect of the two-pair interactions on the thermodynamic potential for various temperatures. This was done by simultaneously solving Eqs.~(\ref{eqrg}) and~(\ref{eqtp}) numerically. We obtain the renormalized Cooper-pair chemical potential at the end of the flow, $\mu_{\Delta, \infty}=\mu_{\Delta}(l\rightarrow \infty)$, which incorporates the pair self-energy effects. The results are shown in Fig.~\ref{figselfen} for the balanced case ($\mu_{\sigma}=\mu$), where the (renormalized) many-body scattering length is plotted, which was introduced in Section \ref{parac}. This scattering length is inversely proportional to the (renormalized) pair chemical potential, namely $1/k_{\rm F} a_{\rm MB}=4\pi \hbar^2\mu_{\Delta}/m k_{\rm F}$ with $k_{\rm F}= (2 m \epsilon_{\rm F})^{1/2}/\hbar$ the Fermi wavevector. In Fig.~\ref{figselfen}, the initial or bare many-body scattering length $a^{\rm B}_{\rm MB}$ and the final or fully renormalized many-body scattering length  $a^{\infty}_{\rm MB}$  are plotted as a function of temperature. We see that $a^{\infty~-1}_{\rm MB}$ is more negative than $a^{\rm B~-1}_{\rm MB}$, which means that the Cooper-pair self-energy, given by $\hbar\Sigma_{\Delta,\infty}=\mu^{\rm B}_{\Delta}-\mu_{\Delta,\infty} $, is positive, caused by the repulsive pair interaction. Alternatively, interpreting the Cooper-pair chemical potential as being proportional to the inverse many-body transition matrix, we have for the considered temperature range that $|a^{\infty}_{\rm MB}|<|a^{\rm B}_{\rm MB}|$, which means that the fluctuation and interaction effects make the gas less strongly interacting. 

\begin{figure}
\begin{center}
\includegraphics[width=1.0\columnwidth]{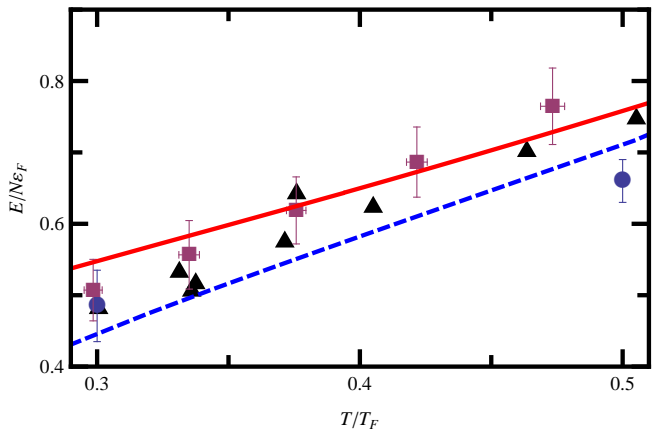}
\caption{\label{figeos} (Color online) Equation of state for the normal phase of a strongly interacting balanced Fermi gas at unitarity in the canonical ensemble. The energy $E$ of the gas is calculated as a function of temperature $T$ with the renormalization group approach (full line) and the Nozi\`{e}res-Schmitt-Rink approach (dashed line). The squares are the experimental results of Harikoshi {\it et al.} \cite{UedaExp}, the triangles of Nascimb\`{e}ne {\it et al.} \cite{SalomonExpNat}, and the dots are the Monte-Carlo results of Burovski {\it et al.} \cite{Burovski}.}
\end{center}
\end{figure}

\begin{figure}
\begin{center}
\includegraphics[width=1.0\columnwidth]{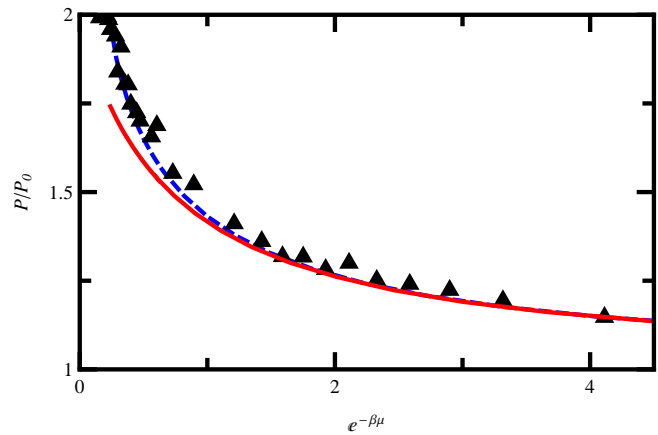}
\caption{\label{figpress} (Color online) Equation of state for the normal phase of a strongly interacting balanced Fermi gas at unitarity in the grand-canonical ensemble. The pressure $P=-\Omega/V$ of the gas is calculated as a function of the inverse fugacity $e^{-\beta \mu}$ with the renormalization group approach (full line) and the Nozi\`{e}res-Schmitt-Rink approach (dashed line). The pressure of the ideal two-component Fermi gas is given by $P_{\rm ig}$. The triangles  are the experimental results of Nascimb\`{e}ne {\it et al.} \cite{SalomonExpNat}.}
\end{center}
\end{figure}

To see whether the selfenergy of the Cooper pairs, or the reduced many-body scattering length, has observable experimental consequences, we calculate the equation of state, which has been accurately measured by Harikoshi {\it et al.} \cite{UedaExp}  and Nascimb\`{e}ne {\it et al.} \cite{SalomonExpNat} for the balanced case. We have that the total thermodynamic potential density is given by
\begin{equation}
\omega (T,\mu_{\sigma})= \frac{\Omega (T,\mu_{\sigma})}{V}=\omega_{\rm ig}(T,\mu_{\sigma})+\omega_{\Delta, \infty}(T,\mu_{\sigma}),
\end{equation}
with $\Omega$ the thermodynamic potential and $\omega_{\Delta, \infty}(T,\mu_{\sigma})= \omega_{\Delta}(l\rightarrow \infty)$. From $\omega$, all other thermodynamic quantities of interest can be obtained by the standard thermodynamic relations. Of particular interest are the particle densities $n_{\sigma} = -\partial \omega/\partial \mu_{\sigma}$ and the energy 
\begin{equation}
E= \Omega + \mu_+ n_+ V + \mu_- n_- V + T S, 
\end{equation}
where $S$ is the entropy, given by $S=-\partial \Omega/\partial T$. The results of our calculations for the balanced case are shown in Fig.~\ref{figeos}, where also the experimental results of Refs.~\cite{UedaExp,SalomonExpNat} are given, as well as the Monte-Carlo results of Ref.~\cite{Burovski}, and the results from NSR theory \cite{NSRpaper,Hu}. It is seen that the inclusion of the Cooper-pair self-energy with the RG theory results in a deviation from the NSR theory that is comparable with the size of the error bar in the experiments and Monte-Carlo calculations. Fig.~\ref{figeos} plots the thermodynamic quantities for the canonical ensemble, while Fig.~\ref{figpress} shows the comparison with the data from Nascimb\`{e}ne {\it et al.} in the grand-canonical ensemble, which gives the most direct comparison with their measurements \cite{SalomonExpNat}. Fig.~\ref{figpress} shows that in the normal state, the agreement with the NSR theory is perfect over a very large temperature range. This is remarkable, since the NSR theory is not exact, and in the strongly-interacting regime deviations might have been expected. With our RG theory we calculate the leading pair self-energy effects beyond the NSR calculation, and find that the effect on the equation of state is quite small, as seen in Fig ~\ref{figpress}. However, the effect is still observable and the agreement with experiments becomes worse. A theoretical explanation for this effect could be that, although we go far beyond the NSR approximation, there are more effects that play a small quantitative role in the comparison with experiments, like for example the renormalization of the effective Cooper-pair mass and three-pair interaction effects. These are interesting topics for further study.

\begin{figure}
\begin{center}
\includegraphics[width=1.0\columnwidth]{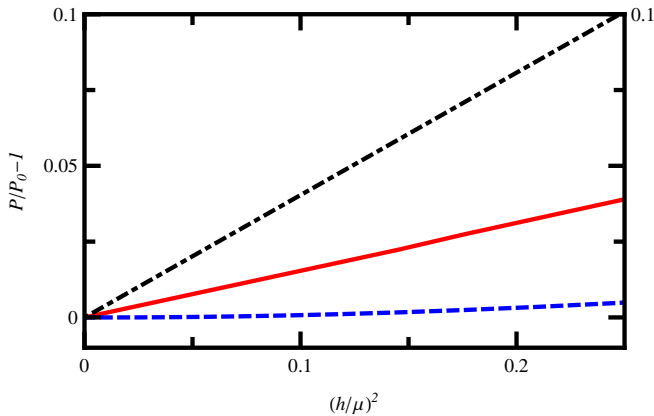}
\caption{\label{figimb} (Color online) Equation of state for the normal phase of a strongly interacting imbalanced Fermi gas at unitarity in the grand-canonical ensemble. The pressure $P=-\omega(T,\mu,h)$ of the gas is calculated at a temperature $T =0.75\mu$  as a function of the normalized chemical potential difference $h/\mu=(\mu_+ - \mu_-)/(\mu_+ + \mu_-)$ with the renormalization group approach (full line), the Nozi\`{e}res-Schmitt-Rink approach (dashed line), and for the ideal Fermi gas (dashed-dotted line). For each curve, the pressure of the imbalanced gas is normalized by the corresponding pressure of the balanced gas $P_{0} = -\omega(T,\mu,0)$. }
\end{center}
\end{figure}

We have also calculated the equation of state for the imbalanced Fermi gas. In Fig.~\ref{figimb}, we show the pressure $P$ as a function of $h/\mu$ for the temperature $T/\mu = 0.75$, where $h=(\mu_+-\mu_-)/2$. At this temperature, the NSR approximation predicts a negative polarization for positive $h$, which is an unphysical result. Our RG theory that treats the interaction effects beyond NSR theory does not have this problem. We see that, as a result, our RG theory, the NSR theory and the mean-field theory give very different results for the pressure of the imbalanced Fermi gas in Fig.~\ref{figimb}. This pressure has recently been measured at zero temperature, where good agreement with Monte-Carlo calculations was obtained \cite{SalomonFL}. The pressure could also be measured above the critical temperature, giving rise to a sensitive test for theories of the imbalanced Fermi gas at nonzero temperatures.

Finally, we briefly discuss the effect of the Cooper-pair interactions on the critical temperature for the balanced Fermi gas. In the unitarity regime, the effective two-pair interaction is repulsive in the normal state. Taking into account the repulsive two-pair interaction will lower the ratios $k_{\rm B}T_{\rm c}/\mu$ and $\epsilon_{\rm F}/\mu$, because physically the repulsive interactions lower the effective chemical potential of the noncondensed Cooper pairs. As a result, the density of noncondensed Cooper pairs is lowered, which decreases the total density and the critical temperature for condensation. Namely, when $G^{\rm B ~ -1}_{\Delta} (0,0) = \mu^{\rm B}_{\Delta} = m/4\pi \hbar^2 a^{\rm B}_{\rm MB}=0$, then both mean-field theory and NSR theory predict a transition to the superfluid state, i.e., the condensation of Cooper pairs. This gives rise to $k_{\rm B}T^{\rm c}_{\rm MF} = 0.66 \mu$, which deviates more than a factor of two from the Monte-Carlo result  $k_{\rm B}T^{\rm c}_{\rm MC} = 0.31 \mu$ \cite{Burovski}. Upon lowering the temperature from $T^{\rm c}_{\rm MF}$, $\mu^{\rm B}_{\Delta}$ becomes positive, but due to the repulsive interactions the renormalized chemical potential is lowered, and at the end of the flow $\mu_{\Delta, \infty}$ might still be negative. Therefore, the critical condition in the presence of Cooper-pair interactions becomes $\mu_{\Delta, \infty}= 0$. With our present theory, this results in $k_{\rm B}T^{\rm c}_{\rm RG} =0.43 \mu $, significantly closer to the Monte-Carlo results than the mean-field and the NSR result. Note that this result can be further improved by performing a RG calculation for the superfluid state. Namely, close to $T^{\rm c}$ we then initially have $\mu_{\Delta} (l) > 0$, and the RG flow starts out in a superfluid state. Due to the repulsive interactions, the effective chemical potential is then again lowered, and the system can flow into the normal state \cite{rgpaper}. This calculation thus requires a superfluid RG, and although progress has been made in this direction using various approximations (such as a low-energy expansion of the Cooper-pair propagator), the full calculation is much more tedious than for the normal phase and is prospect for future research. At this point, we can also explain our choice for the range of the temperature axis in Fig.~\ref{figeos}, since for lower temperatures it would be better to perform the partly superfluid RG.

\section{Discussion and outlook}

To summarize our results, we have calculated the thermodynamic properties of  strongly-interacting Fermi gases. This was achieved by combining two well-established strong-coupling methods, namely the Nozi\`{e}res-Schmitt-Rink theory for the treatment of noncondensed Cooper-pairs together with the Wilsonian renormalization group scheme for treating the pair interaction effects. Our resulting theory incorporates fluctuation effects far beyond the NSR theory, and the obtained equation of state has been compared for the balanced Fermi gas with the experiments of Refs.~\cite{UedaExp,SalomonExpNat}. The difference between the equation of state from our RG theory and NSR theory has about the same size as the error bars in the experiment of Ref.~\cite{UedaExp}.  As a result, taking the theoretical step from the noninteracting NSR theory to the incorporation of pair interaction effects maintains the good agreement with accurate thermodynamic experiments. A detailed comparison between theory and the experiment of Ref.~ \cite{SalomonExpNat} shows that the results from the NSR theory are in perfect agreement over a large temperature range, and the inclusion of self-energy effects makes this agreement somewhat worse. In order to try to understand this result from the theoretical side, the accuracy of the calculation could be further enhanced, namely by calculation of next-to-leading order effects, such as the renormalization of the Cooper-pair mass and the three-pair interaction, which might also have quantitative effects. These are projects for the future. 

The NSR approximation fails in describing the imbalanced Fermi gas, where it leads to unphysical results for small chemical potential differences. In particularly, it predicts a negative polarization $p=(n_{+}-n_{-})/(n_{+}-n_{-})$ for a positive chemical potential difference $(\mu_{+}-\mu_{-})$. However, by including Cooper-pair interaction effects with our RG, this problem is resolved. As a result, we also obtain very different results for the equation of state of an imbalanced Fermi gas compared to NSR theory and mean-field theory. The obtained pressure at nonzero temperatures can be measured in upcoming experiments. Moreover, the NSR theory is not accurate for determining the ratio between the critical temperature and the chemical potential, for which it gives the same result as mean-field theory. Calculating self-energy effects with our RG theory for the balanced Fermi gas results in a critical temperature reduction of more than a factor of 1.5, bringer it closer to Monte-Carlo results. Our method is complementary to a Fermi-liquid theory of the normal state, which is based on single-particle correlations. The Cooper-liquid theory calculates pair-correlation effects, and our method has the advantage that it can in principle be directly generalized to the superfluid state by performing a RG for the pair condensed phase \cite{rgpaper}. Moreover, it can also be extended to the mass-imbalanced case, which has a very rich phase diagram in the unitarity limit \cite{GubbelsMassImb}. These are also projects for the future.

\appendix

\section{Useful relations}

Using contour integration, the Green's function of the Cooper pairs from Eq.~(\ref{eqprop}) can be written in the spectral form
\begin{equation}\label{eqspec}
G_{\Delta}(i\omega_n,k,\mu_{\Delta})=\frac{1}{\pi}\int d\omega \frac{\ {\rm Im}[G_{\Delta}(\omega^{+},k,\mu_{\Delta})]}{\omega-i\omega_n},
\end{equation}
where $\omega^{+} = \omega + i \eta$ with $\eta \downarrow 0$. The imaginary part of the Green's function can be obtained analytically \cite{Falco}. With the spectral representation, we can rewrite Matsubara sums over the pair Green's function as frequency integrals that are convenient for numerical evaluation. For example, we have 
\begin{eqnarray}\label{eqgfint}
&&\frac{1}{\hbar\beta}\sum_n G_{\Delta}(i\omega_n,k,\mu_{\Delta})=\nonumber \\
&&\frac{1}{\pi}\int d\omega N_{\rm B}(\omega) {\rm Im}[G_{\Delta}(\omega^{+},k,\mu_{\Delta})],
\end{eqnarray}
where $n$ is even and $N_{\rm B}(\omega)=1/(e^{\beta \omega}-1)$ is the bosonic distribution function. Moreover, the pair bubble diagram from Eq.~(\ref{eqbubble}) becomes
\begin{eqnarray}\label{eqbubint}
&&  \Pi(k,\mu_{\Delta}) =\frac{1}{\hbar\beta}\sum_n   G_{\Delta}(i\omega_n,k,\mu_{\Delta})^2=\\
&& \frac{2}{\pi}\int d\omega  N_{\rm B}(\omega){\rm Im}[G_{\Delta}(\omega^+,k,\mu_{\Delta})]{\rm Re}[G_{\Delta}(\omega^+,k,\mu_{\Delta})],  \nonumber
\end{eqnarray} 
where we used Eq.~(\ref{eqspec}) and the Kramers-Kronig relation to relate the real and imaginary part of the Cooper-pair Green's function. For the thermodynamic potential density from Eq.~(\ref{eqtp}), we use that \cite{NSRpaper}
\begin{eqnarray}\label{eqtpint}
&&  \frac{1}{\hbar\beta}\sum_n \log[-G^{-1}_{\Delta}(i\omega_n,k,\mu_{\Delta})] = \nonumber\\
&&\frac{1}{\pi} \int d\omega  N_{\rm B}(\omega){\rm Im}(\log[-G^{-1}_{\Delta}(\omega^+,k,\mu_{\Delta})]).
\end{eqnarray}

\emph{Acknowledgements.}--- We thank Munekazu Horikoshi and Sylvain Nascimb{\`{e}}ne for kindly providing us with their experimental data, as well as Evgeni Burovski for sharing their Monte-Carlo data. This work is supported by the
Stichting voor Fundamenteel Onderzoek der Materie (FOM) and the
Nederlandse Organisatie voor Wetenschaplijk Onderzoek (NWO), as well as by the European Community's Seventh Framework
Program ERC-2009-AdG under grant agreement 247142-MolChip.

\bibliographystyle{apsrev}
\bibliography{../library.bib}

\end{document}